\documentclass[submission,copyright,creativecommons]{eptcs}
\providecommand{\event}{SSV 2012} % Name of the event you are submitting to
\usepackage[T1]{fontenc}
\usepackage{textcomp}
\usepackage{
graphicx,
epsfig,
endnotes,
tablists,
booktabs,
algorithm,
algorithmic,
xspace,
booktabs,
url,
amsmath,
amssymb,
amsthm,
stmaryrd,
eucal,
listings,
multirow,
wrapfig,
tikz,
microtype,
balance,
listings,
multicol
}
\usetikzlibrary{shapes,arrows,positioning,snakes}
\usetikzlibrary{calc}

\usetikzlibrary{shapes,automata,arrows,shapes.arrows,shadows,decorations.pathmorphing,fit,backgrounds,plotmarks}

\title{Static Analysis of Lockless Microcontroller C Programs
}
\author{Eva Beckschulze \qquad\qquad Sebastian Biallas \qquad\qquad Stefan Kowalewski
\institute{\\ Embedded Software Laboratory\\
RWTH Aachen University, Germany\\
}
\email{\{lastname\}@embedded.rwth-aachen.de}
}
\def\titlerunning{Static Analysis of Lockless Microcontroller C Programs}
\def\authorrunning{Beckschulze et. al.}

\begin{document}

% maths
\newcommand{\interval}[1]{\left[#1\right]} 
\newcommand{\meet}{\sqcap}
\newcommand{\join}{\sqcup}
\newcommand{\bigmeet}{\bigsqcap}
\newcommand{\bigjoin}{\bigsqcup}
\newcommand{\smaller}{\sqsubseteq}
\newcommand{\wider}{\sqsupseteq}
\newcommand{\set}[1]{\left\{#1\right\}} 
\newcommand{\card}[1]{\left\lvert#1\right\rvert} 
\newcommand{\tuple}[1]{\left<#1\right>}
\newcommand{\ceil}[1]{\left\lceil#1\right\rceil}
\newcommand{\floor}[1]{\left\lfloor#1\right\rfloor}

\newcommand{\magic}{\mathcal{M}}
\newcommand{\rec}{\mathcal{R}}
\newcommand{\signed}[1]{\ensuremath{#1}}
\newcommand{\unsigned}[1]{\ensuremath{\langle #1 \rangle}}

\newcommand{\true}{\ensuremath{\mathit{true}}\xspace}
\newcommand{\false}{\ensuremath{\mathit{false}}\xspace}

\newcommand*{\bdiv}{\mathrel{\mathrm{div}}}
\newcommand*{\sar}{\gg}
\newcommand*{\Z}{\mathbb{Z}}
\newcommand*{\N}{\mathbb{N}}

%\newcommand{\dotcup}{\ensuremath{\mathaccent\cdot\cup}}
%\newcommand{\bigdotcup}{\ensuremath{\mathaccent\cdot\bigcup}}

% modulo
\newcommand{\xfrac}[2]{
	{\raisebox{0.3ex}{$#1$}}
	\hspace{-0.2ex}
	{\raisebox{-0.1ex}{/}}
	\hspace{-0.2ex}
	{\raisebox{-0.3ex}{$#2$}}
}
% x/~
\newcommand{\modtilde}[1]{\xfrac{#1}{\sim}}

\newcommand{\tool}[1]{\textsc{#1}}
\newcommand{\arcadeplc}{\tool{Arcade.PLC}\xspace}
\newcommand{\aplc}{\tool{Arcade.PLC}\xspace}
\newcommand{\mcs}{\tool{[mc]\-square}\xspace}
\newcommand{\murphi}{\tool{Mur$\varphi$}\xspace}
\newcommand{\promela}{\tool{Promela}\xspace}
\newcommand{\spin}{\tool{Spin}\xspace}
\newcommand{\uppaal}{\tool{Uppaal}\xspace}

\newcommand{\code}[1]{\texttt{#1}\xspace}
\newcommand{\todo}[1]{\textbf{(#1)} \marginpar{\bf \textbar}}

% formatting for instruction list identifiers
\newcommand{\IL}[1]{\texttt{#1}}

% Kripke structure
\newcommand\M{\mathcal{M}}

% CTL operators
%\newcommand{\LTL}{LTL\xspace}
%\newcommand{\ptLTL}{ptLTL\xspace}
\newcommand{\LTL}{$\mathsf{LTL}$\xspace}
\newcommand{\ptLTL}{$\mathsf{ptLTL}$\xspace}
\newcommand{\CTL}{$\mathsf{CTL}$\xspace}
\newcommand{\CTLmX}{$\mathsf{CTL}_{-X}$\xspace}
\newcommand{\ACTL}{$\mathsf{\forall CTL}$\xspace}
\newcommand{\CTLop}[1]{\mathit{#1}}
\newcommand{\A}{\CTLop{A}}
\newcommand{\E}{\CTLop{E}}
\newcommand{\G}{\CTLop{G}}
\newcommand{\F}{\CTLop{F}}
\newcommand{\U}{\CTLop{U}}
\newcommand{\X}{\CTLop{X}}
\newcommand{\TT}{\CTLop{TT}}
\newcommand{\FF}{\CTLop{FF}}

\newcommand{\strongInterval}[1]{\left[#1\right)}
\newcommand{\weakInterval}[1]{\left(#1\right)}

% abbrev
\newcommand\eg{e.\,g.\xspace} 
\newcommand\ie{i.\,e.\xspace} 
\newcommand\wrt{w.\,r.\,t.\xspace} 

\newcommand{\acc}{\ensuremath{\mathsf{acc}}}
\newcommand{\var}[1]{\ensuremath{\mathit{#1}}}
\newcommand{\SSA}[2]{\ensuremath{{#1}^{(#2)}}}
\newcommand{\varssa}[2]{\SSA{\var{#1}}{#2}}
\newcommand{\accssa}[1]{\SSA{\acc}{#1}}
\newcommand{\guard}{\mathsf{guard}}
\newcommand{\constraint}[1]{\mathsf{cs}_{#1}}
\newcommand{\constraintsingle}[1]{\constraint{sing}({#1})}
 
\maketitle

\begin{abstract}
Concurrently accessing shared data without locking is usually a subject to race conditions resulting in inconsistent or corrupted data.
However, there are programs operating correctly
%which operate correctly without synchronized accesses
without locking
by exploiting the atomicity of certain operations on a specific hardware.
In this paper, we describe how to precisely analyze lockless microcontroller C programs with interrupts by taking the hardware architecture into account. We evaluate this technique in an octagon-based value range analysis using access-based localization to increase efficiency.
\end{abstract}

%This paper describes a static analysis technique for microcontroller programs written in C that rely on (possible unsynchronized) interrupt invocations.
%To analyze such programs, we implement hardware specific control semantics as well as a memory model 
%in addition to the standard C semantics.
%Crucial to 
%%make 
%our approach 
%%tractable, 
%is a pre-analysis of the interrupts so as to speed up the re-analysis of possible interrupt invocations using localization techniques.
%% octagons?
%The algorithms are implemented in the Spatz framework and evaluated using various microcontroller programs. 
% !TEX root = main.tex
% !TEX spellcheck = en_US
\section{Introduction}
\label{sect:introduction}
Static analysis based on abstract interpretation~\cite{CC77} is a formal method that found its way into practice by several commercial code analysis tools. Proving the absence of run-time errors in microcontroller programs is of particular importance as microcontrollers are often deployed in safety-critical systems.
However, C code analyzers usually do not cope with C extensions and hardware-specific control prevalent in microcontroller programs. 
This control is not only necessary for data input/output but also needed to implement interrupt service routines (ISRs), which allows some form of concurrency and can be used for asynchronous hardware communication and periodic tasks.
Since the control functions of the hardware are often exposed through normal memory accesses, a sound analysis of microcontroller programs has to reflect these registers in its memory model.
On the Atmega16~\cite{atmega16}, for example, it is possible to enable/disable interrupts by a write to the \texttt{SREG} register which is located at the memory address \texttt{0x5F}.

%Microcontrollers have become ubiquitous in practically all electronic devices to control their functionality.
%Since they often control critical functions and their programs are usually hard to update, the correctness of their programs is especially important.
%To analyze such programs, static analysis~\cite{CC77} has been proven to be a promising technique~\cite{citation_needed}.
%Yet, there are certain peculiarities that make the analysis of microcontroller programs challenging, especially when aiming for high precision.
%
%Firstly, microcontroller programs are usually written in a C dialect equipped with extensions that additionally allow for hardware specific control.

Beside these peculiarities in programming microcontrollers, software engineers often rely on additional assumptions outside the scope of standard C semantics on how the compiler will translate a program and on how the microcontroller behaves w.\,r.\,t.~the atomicity of some operations.
For example, they might omit the locking of shared data because an 8 bit read/write is always executed atomically (such algorithms are typically called \emph{lockless}~\cite{ThomasEHart2006a} or lockfree).
This saves program space on the controller as well as execution time but makes a precise analysis on C code level particularly challenging.
In this paper, we deal with atomicity assumptions when analyzing interrupts.
We exploit the characteristics of interrupts w.\,r.\,t. concurrency to design an efficient fixed-point computation.
%As concurrency is a factor increasing costs of the analysis due to the mass of possible execution path efficiency of fixed-point computation is of particular interest. 

%Secondly, analyzing concurrent programs is expensive due to the mass of possible execution paths, which necessitates that (a) interrupts are only analyzed at the necessary locations and (b) the access-set should be restricted to necessary variables to speed up fixed-point detection.
%This is especially important when analyzing over relational domains~\cite{citation_needed}.
%When analyzing at the C level, this immediately raises the question where interrupts can occur

%Finally, the authors of microcontroller programs quite often rely on additional assumptions (outside the scope of standard C semantics) on how the compiler will translate the programs and how the microcontroller behaves w.\,r.\,t. the atomicity of some operations.
%For example, they might omit the locking of shared data because an 8 bit read/write is always executed atomically (such algorithms are typically called \emph{lockless}~\cite{citation_needed}).
%This saves program space on the controller as well as execution time but makes an analysis particularly challenging.
%Yet, due to their small size the are ideal candidates for formal analysis.

%automatic global interrupt disabled in ISRs is an implicit lock. Can we call these programs "`lockless"' anyway?

%advantage of applying access-based localization to analyze interrupt controlled software: Values do not flow arbitrarily through program.

\subsection{Concurrency Induced by Interrupts}
Compared to concurrency implemented by threads concurrency induced by interrupts exhibits some essential differences \cite{RC07}. 
While threads can preempt each other, interrupts can preempt the main program but the main program cannot interrupt an ISR.
An interrupt can only trigger if both the specific interrupt is enabled and interrupts are globally enabled.
Locks to guarantee atomic sections that are not interrupted are, therefore, implemented by disabling interrupts globally.
By default interrupts are disabled in ISRs such that an ISR runs to completion.
Explicitly enabling interrupts in ISRs is allowed but due to the limited stack size an error-prone approach. In this paper, we concentrate on programs without such nested interrupts.
Considering all these specifics of interrupts we can design a more precise analysis of microcontroller C code.

\subsection{Analysis Framework}
We consider interrupts in the context of a data flow analysis evaluating pointers and value ranges based on the octagon abstract domain~\cite{Min06}, in which the relations between variables (memory locations) \texttt{x}, \texttt{y} are expressed as inequalities $\pm \mathtt{x} \pm \mathtt{y} \leq c$, where $c$ is a constant. To consider hardware dependencies, our memory model is augmented with hardware-specific knowledge~\cite{BBSK11} so as to capture, e.\,g., the setting or resetting of interrupt enable bits.
In this paper, we show how to extend this analysis to interrupts by including hardware specifics and taking the C semantics into account.

%To cope with these vexing problems, we harness a combination of different techniques in this paper.
%Our analysis rests on the octagon abstract domain~\cite{Min06}, in which the relations between variables (memory locations) \texttt{x}, \texttt{y} are expressed as inequalities $\pm \mathtt{x} \pm \mathtt{y} \leq c$, where $c$ is a constant.
%Additionally, our memory model is augmented with hardware specific knowledge~\cite{BBSK11} so as to capture, e.\,g., the activation or deactivation of the interrupt enable bits.
%
%%While computational bottle-neck
%This is even more so when taking interrupts into account, which can be executed at every possible program location (or even between statements).
%To alleviate this problem, we use localization \cite{OBY11, BBK12}.
%
%For analyzing lockless algorithms, we distinguish between atomic and non-atomic operations.
%What constitutes an atomic operation (formally introduced later on) is decided 
%, which can be configured on a hardware specific basis.
%Atomic operations (formally introduced later on) 

%\subsection{Localization}

\subsection{Contribution and Outline}
The contribution of this paper is twofold:
(a) We develop a set of rules which lockless C programs must follow to behave predictable under different compilers.
(b) We present a combined analysis of value ranges, pointers and interrupts for lockless microcontroller C programs.
This analysis combines ramifications of the C memory model with understanding of the underlying hardware to allow a sound representation of lockless code.

%\begin{itemize}
%	\item precise analysis of shared data considering the hardware model
%	\item combined analysis of value ranges, pointers and interrupts
%	\item Our analysis is speed up by using a pre-analysis and localization to prevent excessive reanalysis of interrupts.
%\end{itemize}

%Our goal is the precise but sound analysis of shared data in the presence of interrupts. It is not the intention of this paper to check if handling of shared data is implemented correctly.

Our paper is laid out as follows.
First, Sect.~\ref{sect:example} introduces our technique exemplified on a lockless UART driver.
Then, Sect.~\ref{sect:atomicity} details the notion of atomicity we implemented to analyze such programs on a C and hardware-specific level.
Our analysis is described in Sect.~\ref{sect:combined} and is evaluated in Sect.~\ref{sect:case-study}.
The papers ends with a survey of related work in Sect.~\ref{sect:related} and a concluding discussion in Sect.~\ref{sect:conclusion}.
%Given an microcontroller program we improve efficiency of a value range analysis using access-based localization in two ways:
%\begin{itemize}
%	\item Restricting the input context of interrupt service routines to their access sets lowers the number of times ISRs have to be analyzed.
%	\item Additionally, exploiting independencies between main program and interrupt service routines lowers the number of locations where interrupts have to considered.

%\end{itemize} 
% !TEX root = main.tex
% !TEX spellcheck = en-US
\section{Motivating Example}
\label{sect:example}
In this section, we introduce a UART driver that operates without locking shared data by exploiting the atomicity of certain operations on the specific hardware architecture. We discuss different approaches to analyze such a program.

\subsection{Lockless UART driver}
Consider the source code excerpt in Fig.~\ref{code} implementing the receiver of a UART (Universal Asynchronous Receiver Transmitter) driver that is supposed to run on an AVR microcontroller\footnote{This is a slightly modified excerpt of the code found here \url{http://www.mikrocontroller.net/topic/101472\#882716}}.
The driver uses a FIFO \texttt{rx\_buff} to buffer incoming data software-based in addition to the hardware-implemented buffer.
An integer variable \texttt{rx\_in} (\texttt{rx\_out}) is used as an index to denote the position where the next byte is to be stored (where the next byte is to be read).
The function \texttt{getNextPos} is called to increment the index by one or to reset it to 0 when the index is out of bounds.
Reading a byte out of the hardware register called \texttt{UDR} and storing it in the FIFO buffer is performed by \texttt{ISR}, an interrupt service routine that is triggered by hardware.
The function \texttt{getByte} returns the data located at position \texttt{rx\_out}.
We assume that the global interrupt enable bit is set initially and remains set all the time, while the specific interrupt used by the UART is disabled when the buffer is full (line~36) and enabled when there is at least one free position (line~27).
Hence, the functions \texttt{getByte} and \texttt{isEmpty} might be interrupted anywhere in between two operations.
By way of contrast, the ISR always runs to completion due to the automatic global interrupt disable implemented by hardware. As both the main program and the ISR access \texttt{rx\_out} and \texttt{rx\_in} and one of the accesses is a write access, these variables could be subject to a data race.
However, as reading and writing an 8-bit variable on an 8-bit processor architecture is atomic, locking is unnecessary in this case.

	\lstset{mathescape=true, columns =fixed, numbers=left, numberstyle=\tiny, basicstyle=\ttfamily\tiny\mdseries, breaklines=true, numbersep=5pt,tabsize = 2}
	\begin{figure*}
	\hspace*{\dimexpr\fboxsep+\fboxrule+2em}%
	\begin{minipage}{\textwidth}%\dimexpr\textwidth-2\fboxsep-2\fboxrule-5em}
	\begin{multicols}{2}
	\begin{lstlisting}%[gobble=1]
	
	#define vu8(x)  (*(volatile uint8*)&(x))
	
	uint8 rx_buff[RX0_SIZE];
	uint8 rx_in;
	uint8 rx_out;


	uint8 getNextPos(uint8 pos, size){
		pos++;
		if (pos >= size){
			return 0;
		}
		return pos;
	}
	
	
	uint8 isEmpty(){
  	return rx_out == vu8(rx_in);
	}
	
	
	uint8 getByte(){
	  uint8 data;
	  while( isEmpty() );			
	  data = rx_buff[rx_out];		// get byte
	  rx_out = getNextPos(rx_out, RX0_SIZE);
	  URX0_IEN = 1;				// enable RX interrupt
	  return data;
	}
	
	
	ISR( USART0_RX_vect ){
	  uint8 i = rx_in;
		i = getNextPos(i,RX0_SIZE);
	  if( i == rx_out ){			// buffer overflow
	    URX0_IEN = 0;			// disable RX interrupt
	    return;
	  }
	  rx_buff[rx_in] = UDR;
	  rx_in = i;
	}
		
	\end{lstlisting}
	\end{multicols}
	\end{minipage}
	\caption{UART driver}
	\label{code}
	\end{figure*}
	\lstset{basicstyle=\ttfamily\normalsize\mdseries}
	
\subsection{Analyzing Interrupts}
A typical question verified by static analyses is that of all array accesses being within the bounds of the array. This requires a value range analysis of variables to determine possible values for variables used as indices to access an array. Interrupts have to be considered during the analysis in two ways:
First, as calls to ISRs are not visible in the code, ISR code has to be added to the control flow and taken into account appropriately.
For example, this can be done by nondeterministic calls to the ISR between two control flow nodes.
Second, we need to deal with shared variable accesses, i.\,e., an access to a variable in the main program that might be performed incompletely before an interrupt is triggered.
As this may result in corrupted data, care has to be taken for such race conditions to design a sound value range analysis. Next, we discuss the assumptions made during analysis, first in case the analysis is designed for an arbitrary hardware platform and second in case of hardware specifics that can be used to refine the analysis.

\subsubsection{Hardware Agnostic Approach}
Static analyzers for C code usually do not consider any hardware specifics such as interrupts. To analyze microcontroller C programs using generic static analyzers, the user is advised to annotate the program to provide the analyzer with further constraints. Without an appropriate annotation, for example, the analyzer will take the ISR function as dead code since it cannot see the implicit calls by the hardware.
Further annotations are required to deal correctly with atomic sections that can be defined by toggling the interrupt bit in some microcontroller dependent absolute address. The analysis of ISRs is integrated into another analysis by interleaving all expressions outside of atomic sections with non-deterministic calls to ISRs. If the analysis encounters an expression in the main program that accesses a variable that is also accessed by an ISR and one of the accesses is a write access then there is a race condition. Therefore, to be sound, a value range analysis unaware of the hardware architecture has to assume that this variable may take any value within its type bounds. 
We find such race conditions in the UART example in Fig.~\ref{code}:
\begin{itemize}
	\item Read access to \texttt{rx\_in} in line 18, concurrent write access to \texttt{rx\_in} in line 40
	\item Write access to \texttt{rx\_out} in line 26, concurrent read access to \texttt{rx\_out} in line 35
\end{itemize}
Unfortunately, assuming type bounds for these variables that are used as an index to the buffer \texttt{rx\_buffer} results in a presumed array out of bounds access which is spurious. In the next section we discuss how considering a hardware-specific memory model can refine the analysis and avoid this false alarm.

%Programming microcontrollers is intrinsically tied to dealing with the specific hardware.
%Considering hardware features can help to design more precise analyses.

\subsubsection{Considering Hardware Specifics}
Programming microcontrollers is intrinsically tied to dealing with the specific hardware. Taking hardware specifics into account as well when designing static analyses avoids tedious user annotations while increasing precision of analyses. In a hardware-specific memory model, an access to an absolute address (register) is linked to the semantics of this register \cite{BBSK11}. This way, interrupts can be identified and added automatically to the control flow wherever they may occur. We delay the discussion of reducing the number of ISR analyses further to Sect.~\ref{sect:combined}.
%To increase efficiency of the fixed-point computation the number of ISR calls added to the control flow is reduced to a minimum. Analyzing an ISR is necessary between two atomic sections and before and after a shared data access as this access might effect the ISR analysis. In contrast, provided that there are no effects induced by hardware any access to data not accessed in the ISR does not necessitate the previous analysis of the ISR. 
With respect to the shared read/write access mentioned above, hardware considerations enhance the precision of the analysis significantly. Knowing that the target platform is an 8-bit architecture, we conclude that reading or writing to 8-bit variables will always be performed atomically. Thus, \texttt{rx\_in} and \texttt{rx\_out} always have consistent values. To make sure that all possible values are considered when reading \texttt{rx\_in}, it is sufficient to compute a fixed-point over the ISR in advance and propagate it non-deterministically. Similarly, before writing to \texttt{rx\_out}, the analysis of the ISR takes account of the old abstract value of \texttt{rx\_out} while the analysis of the ISR after the write access will consider its new abstract value. Note that it is not the C expression that we assume to be atomic, but the eventual load or store instruction executed by the processor. In the following section, we will detail this notion of atomicity.

%\begin{figure}
%\centering
%			\tikzstyle{line} = [draw, -latex']
%  \tikzstyle{f} = [draw, text centered, text width=6em, minimum height=3cm, node distance = 3cm]
%	\begin{tikzpicture}
%		\tiny{
%		\node (A) at (0,0) {sei();};
%		\node (B) at (2,1) {ISR();};
%		\node (C) at (4,0) {...};
%		\path[line](A) -- (C){};
%		\path (B) edge[loop right](B){};
%		\path[line](A) -- (B){};
%		\path[line](B) -- (C){};
%		\node (D) at (6,0){rx\_out = ...};
%		\node (E) at (8,1){ISR();};
%		\node (F) at (10,0){URX0\_IEN = 1};
%		\node (P) at (4.8,0){...};
%		\path[line] (C) -- (P){};
%		\path[line] (P) -- (D){};
%		\path (E) edge[loop right](E){};
%		\path[line] (D) -- (E){};
%		\path[line] (E) -- (F){};
%		\path[line] (D) -- (F){};
%		}
% \end{tikzpicture}
%\label{cfg}
%\caption{ISR calls added to the control flow graph where it is necessary}
%\end{figure}

% !TEX root = main.tex
\section{Requirements for Lockless Code}
\label{sect:atomicity}
 
\newcommand{\va}{\ensuremath{\mathtt{a}}\xspace}
\newcommand{\vb}{\ensuremath{\mathtt{b}}\xspace}
\newcommand{\vc}{\ensuremath{\mathtt{c}}\xspace}
\newcommand{\vv}{\ensuremath{\mathtt{v}}\xspace}
\newcommand{\vx}{\ensuremath{\mathtt{x}}\xspace}
\newcommand{\vy}{\ensuremath{\mathtt{y}}\xspace}
\newcommand{\vz}{\ensuremath{\mathtt{z}}\xspace}
\newcommand{\V}{\ensuremath{\mathcal{V}}\xspace}
\newcommand{\VX}{\ensuremath{\mathcal{V}_x}\xspace}
\newcommand{\VW}{\ensuremath{\mathcal{V}_w}\xspace}
\newcommand{\VR}{\ensuremath{\mathcal{V}_r}\xspace}

Precisely analyzing shared data in concurrent programs on a high abstraction level such as C code is usually not possible as we are unaware how a compiler translates a C code expression into processor instructions.
On the other hand, stable concurrent programs should not depend on a certain compiler version or compilation options.
In this section, we try to infer basic rules under which lockless C programs can operate in a well-defined manner.
We then enrich these rules by certain hardware specifics:
In the last section, \eg, we argued that writing or reading to an 8-bit variable cannot be interrupted on an 8-bit hardware architecture, which gives rise to a basic form of atomic access.

With the rules derived in this section, we achieve two goals:
First, we can detect program errors (which might manifest themselves depending on compiler specifics) by checking whether a program follows these rules.
Second, we can formulate the foundations of a sound and precise analysis for each program which follows these rules.

\subsection{Atomicity at the Level of C}
\label{sect:atomicity:c}

In the absence of locks, we assume that all data shared between the main function and the ISRs is performed using volatile accesses only (\ie, either by casting the access to a volatile data type or declaring the variable volatile in the first place).
Accessing non-volatile shared data is prone to failure and thus reported, since an optimizing compiler might eliminate seemingly unnecessary reloads and dead stores.
In the following, we call a C expression \emph{competing} if it contains an access to volatile data.
The term competing stresses that competing expressions should be evaluated in a well-defined order.

C compilers are allowed to schedule loads, stores and calculations of expressions in arbitrary order as long as this does not alter the \emph{visible} behavior of the program (as-if rule).
In the presence of volatile accesses, however, at least all volatile objects must be stable at sequence points~\cite[Sect.~5.1.2.3]{c99}.
To illustrate this, consider the shared variables \va, \vb and the statement \va = ++\vb; % with $\va, \vb \in \V$,
which writes twice to shared data between sequence points.
Let us further assume that we have the precondition $\va \leq \vb$.
Although we know that \va and \vb must be stable after this statement, 
the assignment operator forms no sequence point, and a compiler might thus translate this expression into one of the following two pseudo-assembly snippets:

\lstset{mathescape=true, columns =fixed, numbers=left, numberstyle=\scriptsize, basicstyle=\ttfamily\scriptsize\mdseries, breaklines=true, numbersep=5pt,tabsize = 2}
    \begin{minipage}{0.15\textwidth}$\phantom{f}$
    	\end{minipage}
	\begin{minipage}{0.35\textwidth}%\dimexpr\textwidth-2\fboxsep-2\fboxrule-5em}
	\centering
	\begin{lstlisting}
LOAD  temp, b
INC   temp
STORE b, temp
STORE a, temp
	\end{lstlisting}
	\end{minipage}
	    \begin{minipage}{0.15\textwidth}$\phantom{f}$
    	\end{minipage}
	\begin{minipage}{0.35\textwidth}
	\begin{lstlisting}
LOAD  temp, b
INC   temp
STORE a, temp
STORE b, temp
	\end{lstlisting}
	\end{minipage}

In the right snippet, \va is stored before \vb. An ISR invoked between line 3 and line 4 might thus observe that $\va > \vb$, while $\va \le \vb$ is an invariant in the left snippet.
Hence, we want to detect and warn about such statements.

%"At sequence points, volatile objects are stable in the sense that previous accesses are complete and subsequent accesses have not yet occurred."

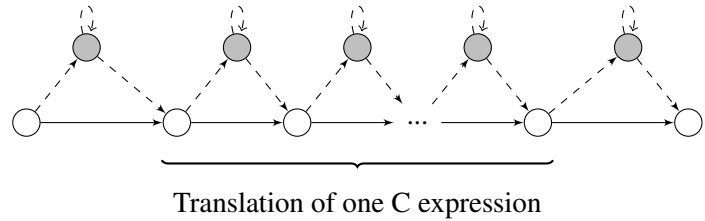
\begin{figure}[t]
\centering
\tikzstyle{line} = [draw, -latex']
\begin{tikzpicture}
\node (A) at (0,0)[circle,draw] {};
\node (B) at (2,0)[circle,draw] {};
\node (C) at (3.6,0)[circle,draw] {};
\node (D) at (5.2,0)[circle] {...};
\node (E) at (6.8,0)[circle,draw] {};
\node (F) at (8.8,0)[circle,draw] {};
\path[](A) -- (C){};
\path[line](A) -- (B){};
\path[line](B) -- (C){};
\path[line](C) -- (D){};
\path[line](D) -- (E){};
\path[line](E) -- (F){};
\node(G) at (0.8,1)[circle,fill=gray!50,draw] {};
\path(G) edge[loop above,dashed](G){};
\path[line, dashed](A) -- (G){};
\path[line, dashed](G) -- (B){};
\node(H) at (2.8,1)[circle,fill=gray!50,draw] {};
\path(H) edge[loop above,dashed](H){};
\path[line, dashed](B) -- (H){};
\path[line, dashed](H) -- (C){};
\node(I) at (4.4,1)[circle,fill=gray!50,draw] {};
\path(I) edge[loop above,dashed](I){};
\path[line, dashed](C) -- (I){};
\path[line, dashed](I) -- (D){};
\node(J) at (6,1)[circle,fill=gray!50,draw]{};
\path(J) edge[loop above,dashed](J){};
\path[line, dashed](D) -- (J){};
\path[line, dashed](J) -- (E){};
\node(K) at (8,1)[circle,fill=gray!50,draw]{};
\path(K) edge[loop above,dashed](K){};
\path[line, dashed](E) -- (K){};
\path[line, dashed](K) -- (F){};
%\draw[snake=zigzag] (1.2,-1) -- (1.2,1.5);
\draw [thick,decoration={brace,mirror,raise=1cm},decorate](1.8,0.5)--(7,0.5);
\node (text) at (4.4,-1.1){Translation of one C expression};
\end{tikzpicture}
\caption{Possible Control Flow on Instruction Level}
\label{model}
\end{figure}

Now, consider  an arbitrary C expression between two sequence points.
Fig.~\ref{model} shows the control flow graph for such an expression assuming that the compiler translated it into a certain sequence of instructions (white nodes), where each instruction might be followed by one or more calls to an ISR (gray nodes).
Between the two sequence points, we do not know how the compiler translated the expression.
This gives rise to two requirements for writing stable lockless code:
\begin{enumerate}
\item The effect of the execution of an ISR between two sequence points must be covered by executing the ISR at the sequence point before or after the expression. 
\item The effect of the execution of an ISR must not depend on the exact instructions generated by the compiler.
\end{enumerate}
The first requirement stems from the fact that the analysis -- as well as the programmer -- can predict the program state only at sequence points, and thus, the ISR behavior should only depend on this predictable state.
The second requirement can be regarded as a corollary of the first: If a compiler creates a different set of instructions (because of, e.\,g., different options), we still want requirement 1 to hold.
We call expressions that fulfill these requirements \emph{well-formed}.

Observe that the well-formedness of expressions depends on two distinct properties:
How much freedom the compiler has to translate an expression (especially scheduling loads and stores) and which atomic primitives are provided by the underlying hardware.
We will defer the discussion of the latter to Sect.~\ref{sect:atomicity:hardware} and for now assume that all loads and stores of shared variables cannot be interrupted by the ISRs.
%We can now investigate certain patterns to characterize well-formed expressions.
%Let $\V=\VR \cup \VW \cup \VX$ denote the set of shared variables.
We will first formally define well-formed expressions inductively:% and then look at some examples. 
%, where \VR is the set of variables that is written in the ISR only, \VW is the set of variables that is only read in the ISR and \VX is written both in the ISR and the main function.
%Let us further assume that $\va, \vb \in \VW$ and $\vx, \vy \in \VR$. We defer the discussion of variables which are written both in the main function and the ISR to Sect.~\ref{TODO}.
%
%We therefore define a 

\newcommand\XOR{\mathbin{\char`\^}}
\begin{enumerate}
\item A constant expression $\mathit{expr} ::= \mathit{const}$ and a variable expression $\mathit{expr} ::= v$ is well-formed.
\item A unary expression $\mathit{expr} ::= \ominus\,\mathit{expr}_1$ with $\ominus \in \set{-, !, \tilde{~} }$ is well-formed iff $\mathit{expr}_1$ is well-formed.
\item Let $\mathit{expr} ::= \mathit{expr}_0 \odot \mathit{expr}_1$ be a binary expression, $\odot \in \{+, -, /, *, \%, \ll, \gg, |, \&, \XOR, <, <=, >, >=,$ $==, !=\}$.
Then, $\mathit{expr}$ is well-formed iff (a) $\mathit{expr}_0$ and $\mathit{expr}_1$ are well-formed and (b) at most one of $\mathit{expr}_0$ and $\mathit{expr}_1$ is competing.
\item Let $\mathit{expr} ::= \mathit{expr}_0 \bowtie \mathit{expr}_1$ be a comma or logic expression, $\bowtie\;\in \{||, \&\&, $``,''$\}$. Since $\bowtie$ forms a sequence point, $\mathit{expr}$ is well-formed iff $\mathit{expr}_0$ and $\mathit{expr}_1$ are well-formed.
\item Let $\mathit{expr} ::= \mathtt{fun}(\mathit{expr}_0, \dots, \mathit{expr}_n)$ be a function call.
Function calls are sequence points, yet the order in which the arguments are evaluated is not defined.
Thus, $\mathit{expr}$ is well-formed if $\mathit{expr}_0, \dots, \mathit{expr}_n$ are well-formed and at most one $\mathit{expr}_i$ is competing.
Additionally, $\mathit{expr}$ is competing iff the body of $\mathtt{fun}$ accesses shared data.
\item An assignment $\mathit{expr} ::= \mathit{lvalue} = \mathit{expr}_1$ is well-formed iff
(a) $\mathit{lvalue}$ is not competing and $\mathit{expr}_1$ is well-formed or (b) $\mathit{expr}_1$ is well-formed and does not write shared data (this does not span into functions which are being called).
\end{enumerate}

Intuitively, a well-formed expression is an expression in which the order of all accesses of shared data is determined by the C standard, \ie, the evaluation of competing expressions is ordered by sequence points.
Additionally, we require that we have at most one write to shared data in a well-formed expression (follows from 6.).
Furthermore, note that the second part of bullet point 5.~forbids delicate cases such as \texttt{a} = \texttt{f()}+\texttt{g()} where \texttt{f} and \texttt{g} access shared data (the order in which \texttt{f} and \texttt{g} are invoked is not defined).
Yet, expressions such as \texttt{a} = \texttt{f()}+1 are well-formed even if \texttt{a} is shared and \texttt{f} is competing.

The latter point deserves a more detailed study:
The assignment operator = does not form a sequence point.
Hence, expressions such as \texttt{a} = \texttt{b} = 0 are not well-formed, since the store to \texttt{a} and \texttt{b} can be performed in arbitrary order.
However,
the compiler must only generate one store to the left-hand-side of an assignment, which depends on the evaluation of the right-hand-side.
Thus, if the right-hand-side contains no writes to shared data, such as in \texttt{a} = \texttt{b}, an assignment is well-formed.
We allow this construct because it is typically used in lockless code.
Alternatively, this construct could be avoided by introducing temporary (unshared) variables.
%\paragraph{Example 1}
%A competing expression that accesses $\vv \in \V$ only once, \ie, \vv is either read or only written to, is well-formed iff the access is performed by an atomic machine instruction (cf. Sect.~\ref{sect:atomicity:hardware}):
%The compiler is required to access the variable only once in this case.
%Thus, each ISR invocation will observe either the old value of \vv or the new, independent of the actual instructions.
%It is therefore sufficient to take ISR invocation before and after the expression into account.

%If a competing expression contains more that one access of shared data, the situation is more delicate:
%\paragraph{Example 2}
%Thus, such expressions are not well-formed.

%
%In general, an expression which reads and writes to shared data is not well-formed.
%Consider the expression ((\texttt{temp} = \vx) + (\va = 1)) which does not specify whether \vx is read first or \va is written first (note that these are assignments, not equalities).
%Still, there are expression such as \va = $\va + 1$ which we would like to regard as well-formed.
%To see this,
%
%The 

%\paragraph{Function Calls}
%Function calls form a sequence point~\cite[Sect.~6.5.2.2, sentence 10]{c99}.
%We still have to account for certain specifics:
%(a) The order in which the arguments are evaluated is not specified, so we can allow only one shared variable here.
%(b) We can have multiple functions calls between two sequence points.
%In this case, the order in which the functions are called is not specified.

The design rules that we derive to achieve robustness of lockless programs against compiler reordering are simple. Function call expressions should be either full expressions or should adhere to bullet point 5. Further, the programmer is advised to avoid subexpression with side effects (accessing a volatile object is a side effect, too). By splitting up complex expression in several simple ones without side effects in subexpression or by at least encapsulating them in simple assignments the desired order of evaluation is made explicit. These are well-known design rules that usually aim at well-defined (compiler-independent) behavior of single-threaded code. However, it also contributes to well-formed expressions which matter in lockless concurrent programs only.

Finally, we require the compiler to behave \emph{sensible}.
That is, in essence, that loads and stores to volatile data are translated to elementary load and store operations (one instruction where possible).
Note that this requirement is, to the best of our knowledge, fulfilled by all compilers used in industry and is exactly what a programmer expects -- and exploits -- when writing lockless code.
In the next section, we will connect the general requirements of C to the specific offerings of the hardware to create a concise memory model.

%Our analysis rests on an intermediate representation of the C code.
%Each node is a three address code of the form $\mathtt{a} = \mathtt{b} \odot \mathtt{c}$, where $\odot \in \{+,-,=,*,\}$

\subsection{Atomicity on the Hardware Level}
\label{sect:atomicity:hardware}

Crucial for our approach is a hardware model that reflects atomicity at the assembly level.
In the last section, we assumed that each elementary load and store of shared data can be performed atomically, \ie, it cannot be interrupted. % by an ISR.
Depending on the type and size of the data object, this is in general not the case.
However, accessing data of the size of general purpose registers can usually be performed by one instruction, which then is not interruptible.
Loading and storing data of different sizes, on the other hand, is usually performed by a sequence of instructions.
Thus, an ISR might interrupt this sequence and read or modify corrupted data.
Our analysis allows to configure the size of atomic data types beforehand
and aims to detect cases where atomic accesses are required but not possible.
%Beside these loads and stores, other atomic instructions (or sequences of instructions) are given by:

Additionally, some processors provide atomic primitives such as compare-and-swap as one instruction. These instructions can only be accessed using compiler specific functions, which are usually built-in or provided as inline assembly.
Our analysis can simulate these instructions atomically if the functions provided by the compiler are  annotated appropriately.
%\end{itemize}
With this structure in place, we can now describe our analysis in the context of interrupts and lockless code.
\section{Designing Analyses Considering Interrupts}
\label{sect:combined} 
Analyzing ISRs and the sound handling of shared data requires adaptations of the existing data flow analysis. In the following, we first describe the original analysis and subsequently discuss how to integrate ISR analysis and shared data handling. 

\subsection{Original Analysis}
The original analysis is a combined analysis of pointers and value range analysis based on octagons~\cite{Min06}. It is a fixed-point computation iterating over the nodes in the control flow graph (CFG) and propagating the computed results along control flow edges until old and new results coincide. The analysis is flow and context sensitive, \ie, it determines an abstract value for each node in the CFG and distinguishes different calling contexts for function calls. The octagon analysis evaluates nodes concerning value ranges while an address taking node is handled by the pointer analysis. 

In order to increase efficiency of the octagon based analysis, we apply access-based localization, a technique that reduces the abstract state that is propagated when analyzing a function to the subset of the abstract state that is actually accessed in this function or in subroutines~\cite{BBK12, OBY11}. The set of accessed variables in each function is over-approximated by a pre-analysis based on a flow-insensitive pointer analysis.

\subsection{Invoking Interrupts in Fixed-Point Computation}
For efficiency reasons, ISRs shall only be analyzed if necessary. Therefore, we make use of our memory model \cite{BBSK11} in order to find accesses to absolute addresses that correspond to setting or clearing the global or individual interrupt bits. The current status of the interrupt bits is added to the abstract state propagated along the CFG.
The na{\"i}ve approach would be to trigger ISR analysis in between two arbitrary nodes where the interrupt is enabled.
However, the execution of an ISR does not necessarily affect all subsequent operations of the main program. Precisely stated, an ISR affects an operation in the main program if and only if the operation accesses data that might be written by the ISR. In order to incorporate such dependencies ISRs have to be analyzed before analyzing a shared data access. In our analysis, we trigger ISR analysis after each node that enables interrupts (globally or individually). This way we also ensure that ISRs are analyzed at least once between two atomic sections. 

\lstset{mathescape=true, columns =fixed, numbers=left, numberstyle=\tiny, basicstyle=\ttfamily\tiny\mdseries, breaklines=true, numbersep=5pt,tabsize = 2}
	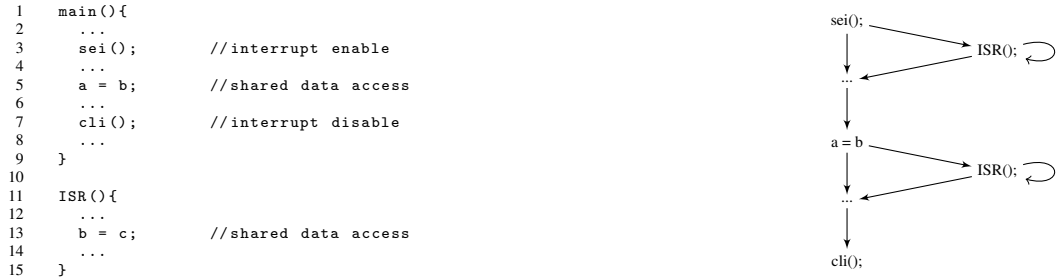
\begin{figure*}
	\hspace*{\dimexpr\fboxsep+\fboxrule+2em}%
	\begin{minipage}{0.5\textwidth}%\dimexpr\textwidth-2\fboxsep-2\fboxrule-5em}
	\begin{center}
	\begin{lstlisting}
	main(){
		...
		sei();       //interrupt enable
		...
		a = b;       //shared data access
		...
		cli();       //interrupt disable
		...
	}
	
	ISR(){
		...
		b = c;       //shared data access
		...
	}
	\end{lstlisting}
	\end{center}
	\end{minipage}
	\begin{minipage}{0.5\textwidth}
	\centering
			\tikzstyle{line} = [draw, -latex']
  \tikzstyle{f} = [draw, text centered, text width=6em, minimum height=3cm, node distance = 3cm]
	\begin{tikzpicture}
		\tiny{
		\node (A) at (2,5) {sei();};
		\node (B) at (2,4.2) {...};
		\path[line](A) -- (B){};
		\node (C) at (2,3.4) {a = b};
		\path[line](B) -- (C){};
		\node (D) at (2,2.6) {...};
		\path[line](C) -- (D){};
		\node (E) at (2,1.8) {cli();};
		\path[line](D) -- (E){};
		\node (F) at (4,4.6){ISR();};
		\path[line](A) -- (F){};
		\path[line](F) -- (B){};
		\path (F) edge[loop right](F){};
		\node (G) at (4,3){ISR();};
		\path[line](C) -- (G){};
		\path[line](G) -- (D){};
		\path (G) edge[loop right](G){};
		}
 \end{tikzpicture}
	\end{minipage}
	\caption{Code example (left) and its control flow graph (right) enhanced with necessary non-deterministic calls to the ISR that need to be considered in the analysis}
	\label{GenericCode}
	\end{figure*}
	\lstset{basicstyle=\ttfamily\normalsize\mdseries}
	
With respect to the control flow, the ISR analysis corresponds to an extra node calling the ISRs added between an interrupt-enabling node and its successor node while the direct connection between these nodes is kept to consider the case that no interrupt is triggered (cf. Fig~\ref{GenericCode}). 
The analysis applies the join operation to the abstract states of both branches.
We compute a fixed-point over all ISRs that might be triggered considering every possible order and frequency of interrupts in order to over-approximate the number of possibly occurring interrupts. Indeed, on an AVR microcontroller each ISR execution is followed by at least one main program instruction. However, the exact number of possible ISR executions cannot be determined on C level, as we are unaware how the compiler translates a given code sequence into instructions.

Still, analyzing interrupts at one location between two atomic sections is not sufficient if shared data is written in the main program and read by the ISR. Therefore, we trigger another analysis of ISRs after analyzing a shared access node (cf. Fig~\ref{GenericCode}). As previously noted, interrupts in the analysis are considered non deterministic but possibly occurring infinitely often. Note that analyzing ISRs both before and after a shared access node is sufficient if the requirements in Sect.~\ref{sect:atomicity:c} are fulfilled.

Finally, to speed up the analysis, we apply access-based localization as explained above also to ISRs. For this purpose, we arrange the pre-analysis collecting accessed variables to analyze ISRs as well. Additionally, we subdivide accessed variables into read and written variables in order to be able to distinguish the different cases of shared access.

% While we delay the discussion of handling both suspicious and well-defined cases of shared data to the next section, we observe that writing shared data in the main program might affect ISR analysis. Consequently, a shared data access between two atomic sections necessitates subsequent ISR analyses as well. 
%In order to avoid unnecessary paths in the control flow graph we reduce ISR invocations to a minimum. Consider an arbitrary shared access to one or several variables as visualized in Fig.~\ref{}. Obviously, an ISR might effect the analysis of the shared access node in main requiring ISR analysis in advance. For simplicity we assume that ISR analysis is required at least once before entering the next atomic section. Therefore, we trigger interrupt analysis immediately after the node that enables interrupts. Although the number of times an interrupt can be triggered in between two arbitrary statement is limited by the fact that an AVR microcontroller executes at least one instruction of the main program in between two ISR execution, the exact number cannot be determined on C code level. We provide this by simply computing a fixed-point over the ISRs.

\subsection{Sound Analysis of Shared Data}
%\textbf{SB suggestion}: If an instruction is well-formed (cp.~Sect.~\ref{sect:atomicity:c}) then it is sufficient to analyze ISR before \& after the instruction.
%If the instruction is not well-formed: type bound for all vars + issue warning.
Shared data handling depends on whether a full expression (cf. \cite[Sect. 6.8]{c99}) is well-formed. As our control flow nodes often represent subexpressions of a full expression, we add to each node whether the corresponding full expression is well-formed.
%The concrete case of shared data decides on how to analyze the shared data node.
Table~\ref{cases} reviews the cases of shared accesses and indicates the behavior of the analysis. We omitted the case where both the main program and an ISR only read the same variable as it does not raise any problem.

The first row of the table shows the case that the shared accesses are atomic, the main program writes(reads) and an ISR read (writes) the shared variables and the corresponding full expression is well-formed. In this case the octagon analysis is performed as usual. All issues of concurrency are in this case considered by triggering the ISR analysis somewhere before and immediately after this node.
In the second row we consider a well-formed full expression that includes an atomic write of a variable that is also written in an ISR. Though it does not cause corrupted data, one write may immediately overwrite the other one. As this might be unintended by the programmer, we issue a warning that data loss might occur.

The third row considers the case that the analysis encounters a full expression where all shared accesses are atomic but the expression is not well-formed. In this case the analysis sets all shared variables to type bounds thereby overapproximating any possible order of execution. Note that these overapproximations are also propagated to the subsequent ISR analysis. We issue a warning that this expression is unspecified behavior. The case that the access is non-atomic (row~4) is handled by the analysis the same way. Here, we issue the warning that a non-atomic access might result in inconsistent or corrupted data.

\begin{table}
\centering
\begin{tabular}{cccc}
\toprule
	atomic  & access & well-formed & \multirow{2}{*}{analysis behavior} \\
	access & type & expression & \\
	\midrule
	 yes & r/w & yes &  no special behavior \\
	 yes & w/w & yes &  issue a warning \\		
	 yes & * & no & set shared variables to type bounds, issue a warning \\	
	 no  & * & * & set shared variables to type bounds, issue a warning \\	
\bottomrule
\end{tabular}
\caption{Cases of shared access}
\label{cases}
\end{table}

\section{Case Studies}
\label{sect:case-study}
Our implementation is written in \textsc{Java} and builds on the Eclipse Framework.
We used it to analyze several microcontroller programs written for the AVR ATmega~16 microcontroller unit.
For this processor our analysis assumes that only 8 bit memory accesses are performed atomically.

The results of the case study are presented in Table~\ref{tbl:casestudy}.
For each checked program, we provide the lines of code, number of global variables, time for analysis, number of times we have to check the ISRs and the number of (spurious) warnings. 
The analyzer was able to prove the absence of array out of bounds accesses in the UART buffer.
Yet, we still had a spurious warning, since this program writes to the data buffer without locking from both the ISR and the main function.
The correctness of such an operation has to be checked manually.
In the RGB-LED program, we  found an unlocked shared access to a 16-bit variable, which was a legitimate warning.
Finally, the Traffic Light program, controlling an intersection with two traffic lights, could be checked without triggering a warning. 
%Our analysis always omits a warning in the following cases:
%\begin{itemize}
%\item A shared resource is accessed without \texttt{volatile} keyword (either at the declaration or the access site).
%\item Shared data is accessed without locking except in the allowed cases in Sect.~\ref{sect:atomicity}.
%\end{itemize}
%Although our analysis continues in these cases by assuming undefined data for the affected variables, we still omit a warning.
%We do this, on the one hand, since such cases almost always indicate a problem, which can easily be fixed.
%On the other hand, these cases exhibit undefined behavior according to the C standard.
%This entails that even this code works it is highly fragile and prone to break by changing the compiler or compiler switches.

\begin{table}
\centering
\begin{tabular}{lrrrrrr}
\toprule
\textbf{Program} & \textbf{Loc} & \# \textbf{global vars} & \textbf{Time} &  \textbf{\# of ISR analyses} & \textbf{\# Warnings} & \textbf{\# Warnings}\\
 & &  & &   & (legitimate) & (spurious)\\
\midrule
UART buffer & 175 & 32  & 7.3 s & 136 & 0 & 1\\
RGB-LED & 95 & 22 & 0.8 s & 27 & 1 & 0\\
Traffic Light & 68 & 5 &  0.1 s & 30 & 0& 0\\
\bottomrule
\end{tabular}
\caption{Results of the case study}
\label{tbl:casestudy}
\end{table}

% !TEX root = main.tex
\section{Related Work}
\label{sect:related}

Traditionally, model checking has been used to verify concurrent programs such as in \cite{God96} where partial order reduction is used to increase efficiency.
%The strive for efficiency gave rise to the techniques of partial order reductions, ... .
Schlich et al.~\cite{SNBB09} implement this technique for model checking embedded software on the assembly level.
Atig et al.~\cite{ABBM10} describe how to model check in the presents of a weak memory model, which corresponds to the lockless programs described in this paper.

Regehr and Cooprider~\cite{RC07} describe how to map microcontroller programs (interrupt-driven code) to thread-driven code. 
In particular, they point out the differences between threads and interrupts and show how to exploit existing techniques for verification tools for multithreaded code to verify interrupt-driven embedded software. Cooprider~\cite{Coo08} describes how to increase efficiency by only analyzing ISRs at certain locations.
His approach, however, is restricted to properly locked programs.

Pratikakis et al. describe in \cite{PFH11} the LOCKSMITH tool which can check automatically for proper locking of shared data in an abstract interpretation framework.
We focus on the verification of microcontroller programs even in the absence of locks.

Recently, Min{\'e} presented an extensive work~\cite{Min12} on analyzing embedded real-time parallel C programs based on abstract interpretation.
He defines the semantics of expressions based on interference, \ie, whenever a thread reads a variable \texttt{x}, all abstract values another thread might set \texttt{x} to are also considered. These interference sets are non-relational and flow-insensitive information while we interchange relational and flow-sensitive information between the main program and ISRs. Due to the flow-insensitivity dealing with the order of execution of C expressions is superfluous in Min{\'e}'s approach, while it is essential in our approach.
% He deals with the uncertainty of how a compiler translates a C expression by showing that the used semantics are invariant under certain program transformations such as reordering assignments and expression simplification.
Additionally, his approach differs from ours in considering all primitive statements to be atomic independent of types and the underlying hardware. This way shared data access is not handled correctly in case of incomplete assignments.

%TODO: Should we discuss verification of assembly?

In \cite{FHST09} Fehnker at al. extend the generic (unsound) static analyzer Goanna to detect microcontroller specific bugs. In their approach, the CFG is labeled with occurrences of syntactic constructs of interest, while the desired behavior is put into a CTL formula that can be checked by model checking techniques. Their work focuses on integrating hardware information to specify and check simple rules that should be followed when programming the specific microcontroller. Instead, this paper advocates the use of a hardware-specific memory model to enhance precision of data flow analyses and to avoid false alarms.

Finally, there is an ongoing effort to formalize memory models for existing languages, which recently cumulated in a memory model for the new C++ standard~\cite{BA08}.
Yet, programmers of microcontroller software still rely on non-strictly defined semantics and ``sensible'' compiler behavior mixed with knowledge about the underlying hardware.
Our approach aims to implement these assumptions, which are sometimes quite subtle, in a verification framework.
 
% !TEX root = main.tex

\section{Concluding Discussion}
\label{sect:conclusion}

This paper advocates a static analysis for lockless microcontoller C programs combining different techniques to make the approach precise as well as tractable.
To achieve precision for such programs, it is necessary to deal both with C semantics and hardware specifics.
Our memory model reflects what the user can (and will) expect from the compiler on the one side and what atomic primitives the hardware can provide on the other side.
Using the derived rules for well-formed expressions, we can detect latent bugs that would manifest themselves only in certain circumstances (such as different compiler options) and we can detect bugs that result from improper communication between the main program and the ISR using our fine-grained value analysis.

Still -- as Meyers and Alexandrescu~\cite{MA04} point out -- thread-unaware languages such as C pose inherent difficulties to write thread-aware code.
It makes proper (manual) synchronization exceptionally hard since all corner cases of the languages have to be considered.
Additionally, as we have shown in this paper, it poses obstacles for the program analysis, since the order of certain operation is often unclear.
In such cases, either imprecise non-relational analyses have to be deployed or a careful analysis of all C expressions is necessary.

%The interplay of these two realms allow us to analyze lockless programs, whereas the localization techniques were used to speed up the process.

%This paper advocates a static analysis for analysis lockless microcontoller programs which combines different techniques to make the approach precise as well as tractable.
%To achieve precision, we use a relational domain which operates on a hardware-specific memory model.
%This memory model reflects what the user can (and will) expect from the compiler on the one side and what atomic primitives the hardware can provide on the other side.
%The interplay of these two realms allow us to analyze lockless programs, whereas the localization techniques were used to speed up the process.

%Future work includes,  

\section*{Acknowledgements}
This work was supported, in part, by the DFG research training group 1298 \emph{Algorithmic Synthesis of Reactive and 
Discrete-Continuous Systems} and by the DFG \emph{Cluster of Excellence on Ultra-high Speed Information and Communication}, German 
Research Foundation grant DFG EXC 89.
Further, the work of Sebastian Biallas was supported by the DFG.
We thank Volker Kamin for sharing his thoughts on the ideas described in this paper, and the anonymous referees for their helpful comments.

\bibliographystyle{eptcs}
\bibliography{ssv}
\end{document}